# Cultivating the Next Generation: Outcomes from a Learning Assistant Program in Engineering

Ying Cao[1], Christina Smith[2], Ben Lutz[1], and Milo Koretskey[1]

*1. Oregon State University*
*2. Brown University*

## Introduction

A growing tension in higher education in science, technology, engineering, and mathematics (STEM) disciplines is the need to produce a greater number of STEM graduates [1] while maintaining learning effectiveness in the resulting large-enrollment STEM courses. One way to mitigate this tension is to create active learning environments and thus increase student engagement and improve learning [2] - [4]. To realize and enhance active learning classes, undergraduate learning assistants (LAs) appear as catalysts. LAs are undergraduate students who have typically completed the particular course and return to assist with its instruction. Practicing LAs increase interactive engagement of the students in active learning classes by providing near-peer help. The processes of facilitating student learning are also construed as a learning experience of LAs themselves.

The generalized LA program developed by the *Learning Assistant Alliance* has three core elements [5]. First, LAs receive professional development in pedagogy during their first academic term as an LA. Second, they work regularly with the course instructor as a member of the instructional team to better understand the content that they will deliver in class. Third, they facilitate active learning in classes of near peers, and reflect on their learning and practice in writing. LAs have become widely used in science courses at many universities and there is research evidence that the programs effectively enhance the success of the students in LA-facilitated courses and of the LAs themselves [6], [7]. To date, the implementation and research about engineering LA programs is sparse.

At a large public university, we identified specific logistical barriers and educational goals in the College of Engineering and adapted the LA Program developed in the College of Science to meet the needs in engineering. In this paper, we describe characteristics and rationale of the adapted Engineering LA Program and study the desired learning outcomes of the engineering LAs. We ask the following research questions.
(1) What are the ways that LAs perceive that the Engineering LA Program helps them learn knowledge and skills that are useful in engineering practice?
(2) How do the LAs' perceptions of learning align with their description of their roles in the course and the overall LA experience?

## Background

The LA Program at Oregon State University began in the College of Science in 2014. It was initiated in the Biology department as part of a larger organizational change initiative *Enhancing STEM Education at Oregon State University* (ESTEME@OSU) to implement evidence-based instructional practices across STEM departments in engineering and science. Integration of pedagogically- and content-prepared LAs into the course instructional team can provide needed support for evidence-based instructional practices. As of fall 2017, the LA Program had spread to include courses in five out of seven departments in the College of Science and four out of six departments in the College of Engineering.

The ESTEME@OSU uses communities of practice (CoPs) as the primary mechanism for implementation and scaling the use of the evidence-based instructional practices. CoPs allow faculty who have been independently developing and implementing similar innovative instructional practices to regularize across departments. The community of practice supports further development – allowing innovators to borrow from one another and to collectively address problems they cannot solve independently. The CoPs facilitate evolving relationships amongst members developed around things that matter. Our approach is based on the premise that in the inclusion of three interacting elements - (i) using community-agreed upon evidence-based instructional practices; (ii) while working to increase scale, and (iii) learning about what other units are doing and how they are doing it through CoPs - we have components for emergent organizational change.

Through the CoPs, we identified a persistent problem of practice for engineering LAs. They did not have room in their curricular plans to add the 2-credit LA pedagogy course as dictated in the College of Science LA Model. Therefore, we adapted the program to provide a similar experience using a non-class workshop delivery for pedagogical development. The workshop contained 8-hours of contact time, including a 4-hour pre-term meeting, and two 2-hour meetings during the term. The LAs also completed 10 weekly online reflections. Workshop topics included learning theories and pedagogical methods. They also situated the theories and pedagogy within the context that LAs would be working. The materials were adapted from the LA resources available at the *Learning Assistant Alliance* website [5]. The online weekly reflections were set up in the *AIChE Concept Warehouse* [8]. Each week, LAs received a prompt asking them to read a short article about learning and pedagogy and relate that to their teaching experiences through a 250-word written reflection.

For recruitment of faculty to include LAs in their course instructional team, we targeted large-enrollment classes (over 100 students), but did not exclude other courses with enthusiastic instructors. We specifically targeted introductory courses that had a history of hiring undergraduate students to facilitate laboratories and recitations. The LA Program added the pedagogy elements (both the workshop and the online reflection) and, in some cases, shifted to more structured, regular meetings with the instructional team; thus, we used existing structures to transition from "undergraduate TAs" to LAs. There were also courses that never used undergraduate TAs before but started to include undergraduate LAs in class to work alongside graduate TAs and help facilitate active learning. All LAs in engineering received a stipend for their pedagogical development (the workshop and online reflection) in addition to hourly payment for their role in course instruction.

This study collected data from the fall quarter 2016 LA cohort, which was the first time the LA Program we formally used in engineering. Our goal was to provide the engineering LAs the same level of professional development as the College of Science LAs, but to accommodate the engineering students' already high credit loads (and therefore resistance to pay for course credits).

## Theoretical Framework

We frame this study as to capture the socio-technical aspects of the landscape of engineering practice [9], [10] that emerged in LAs' description of their experience and learning. By socio-technical we mean that engineering relies on social processes at different levels of interactions between people and is centered on technical work. Engineering can involve working with and influencing other people, and, in addition, there is often interplay between the ways engineers engage in social processes and the nature of the technical work that results [11]. Research on engineering practice has revealed interwoven aspects of engineering knowledge and skills such as technical knowledge, communication of technical knowledge, socio-technical coordination, team cooperation, and managerial skills, e.g., [12] - [14].

In this study we focus on the socio-technical development of the LAs themselves, as part of investigating how the LA program influences the learning outcomes of undergraduate engineering students involved in this program (including the students in the active learning classes that LAs facilitate and the LAs). The nature of engineering LA practice—facilitating peer students' learning of a subject as they complete a design, laboratory, or problem-solving task—can be considered from the perspective of how it helps develop LAs' knowledge and skills that will be useful for engineering practice. Through analysis of the LAs' interview responses, we hope to learn how LAs' perceptions of their learning fits the broader spectrum of socio-technical knowledge and skills and what factors may have influenced it. The knowledge resulting from this study can help us modify the LA Program to improve the learning outcomes as well as inform others who are interested in using LAs in engineering.

## Method

### *Research design*

The study reported in this paper primarily analyzed interview data from 11 engineering LAs to answer the research questions about the LA's learning experiences and their socio-technical knowledge and skill development. It is part of a broader research study to describe and evaluate the uptake of the LA Program at Oregon State University (the Overall LA Study). In turn, this Overall LA Study is situated as a case study in the overarching organizational change initiative at Oregon State University, described earlier in the background section.

We designed the Overall LA Study in a way to allow us to collect data from multiple sources to triangulate our analyses [15]. Data sources from Fall 2016 are shown in Table 1 and include LA and instructor interviews; a pre- and post- term LA survey; the LAs online weekly reflections; observations of the LAs in action (at teaching sessions, meetings, and workshop sessions); and LA Program development documentation (such as plans and meeting notes). The Institutional Review Board approved this study and all participants provided informed consent.

**Table 1.** Data collection summary from Fall 2016

| # of LAs interviewed | # of instructors interviewed | # of survey responses | # of reflection responses | # of sessions observed | # of course artifacts |
|---|---|---|---|---|---|
| 11 | 4 | 13 | 52 | 26 | 28 |

Note: # represents *Number*.

*Context*

Table 2 shows the engineering courses in Fall 2016 that had LAs, the context LAs participated in teaching, the number of students enrolled in the LA-facilitated courses, the number of LAs instructing in specific group of classes, and the number of LAs that attended the LA pedagogy workshop.

**Table 2.** Engineering LA course implementation data for Fall 2016

| LA-facilitated Courses | Department | LA context | Students enrolled | # LAs in class | # LAs in workshop |
|---|---|---|---|---|---|
| Material Balances (CBEE 211) | Chemical, Biological and Environmental Engineering | Studio | 259 | 5 | 5 |
| Introduction to MIME (MIME 101) Engineering Computing | Mechanical, Industrial, and Manufacturing Engineering | Recitation Labs | 417 | 19 | 16 |
| Electrical Fundamentals Digital Logic Design Laboratory Introduction to ECE (ECE 111) | Electrical Engineering and Computer Science | Labs Office hours | 523 | 20 | 16 |
| Engineering Orientation (ENGR 111, ENGR199) | Engineering | Recitation | 251 | 17 | 13 |
| Totals | | | 1,450 | 61 | 50 |

*Data collection*

In this present study, we focus analysis on the LA interview data; however, the other sources in Table 1 were used to triangulate findings and consider implications. After the term ended, all the LAs were invited to participate in a one-on-one, 60-minute interview with the first author. Eleven LAs consented and completed the interview. All the interviews were audio recorded. Of the 11 interviewed LAs, three facilitated studios in a chemical, biological, and environmental engineering course, four taught recitations (two in an introductory general engineering course and two in an introductory mechanical, industrial, and manufacturing engineering), and four taught labs in an electrical engineering course. One was an Asian American male; one was a Black international male; four were White American females; and five were White American males.

The LAs were asked about their experience the past term. They were asked about *facts* (e.g. "what courses did you teach" or "what classroom activities did you participate in?"); about their understanding of their *role* as an LA; about their understanding of the *knowledge and skills* LAs should have; about their LA *experience* (such as a story, the successes, and the difficulties); and about their *perception of learning* (what they have learned from the experience and how that learning outcomes can be transferred in other situations). The interview protocol was semi-structured and the interviewer decided whether and to what extent to ask *probing* question (e.g., "why do you think this story is special? and "how did you deal with the difficulties?")

*Data analysis*

We used an emerging coding approach to parse the transcribed interview recordings. We used LAs' answers to the facts question*s* to learn the context of LAs' practice and did not perform coding on them. We also did not code on the question*s* that we did not ask in the same way to every LA. The interview questions that we have included in the coding process are shown in Table 3.

**Table 3.** Questions coded in this study

| Category | Questions |
| --- | --- |
| Role | Can you describe your understanding of the role of a learning assistant? |
| Knowledge and skills | What knowledge and skills do you think are needed to be an LA? |
| Experience | Can you share with us a story that comes to your mind about your LA experience that particularly influenced you? |
| | What are the successes that you have had as an LA? |
| | Have you encountered any difficulties? |
| Perception of learning | Has working as an LA changed your own thinking about teaching and learning? |
| | How do you think your experience as an LA will help you as a student? Why? |
| | How do you think your experience as an LA will help you in the future? Why? |
| | What do you wish you would have known early on in your role as an LA? |
| | What will you do differently next time you work as an LA? |
| | What do you think you have learned from the LA experience that you wouldn't be able to learn without this experience? |

To answer our first research question: *What are the ways that LAs perceive that the Engineering*

*LA Program helps them learn knowledge and skills that are useful in engineering practice?* we identified when the LAs described what they have learned from the experience and assigned them codes that categorize the kinds of learning they described (e.g., a better understanding of content knowledge, better communicating with people, facilitating group work). Especially, we focused on a group of interview questions that were designed to elicit LAs *perception of learning*: what they have learned from the experience and how those learning outcomes can be transferred in other situations. These questions included the LAs' perception of the changes in their own thinking about teaching and learning; things they wish would have known earlier; things they would do differently next time; learning that would help them as students and in future; and things they wouldn't be able to learn without the LA experience. We obtained set of emergent categories as shown in Table 3.

**Table 4.** Emergent categories of LA learning.

| Categories of learning | When LAs mentioned learning about |
|---|---|
| Understanding content | course content<br>disciplinary concepts<br>lab equipment |
| Communicating disciplinary knowledge | explaining concepts or problems<br>guiding student thinking<br>asking strategic questions<br>answering student questions<br>helping students understand course content |
| Understanding others | other ways of thinking or approaches<br>other perspectives or roles (e.g., professor vs. student) |
| Team work | facilitating group work<br>participating in group work |
| General communication | communicating with individuals<br>public speaking<br>dealing with complex social situations |

In these code categories, *understanding content* means the LAs' own understanding of the course content. *Communicating disciplinary knowledge* means that the LAs and students were talking about technical questions and concepts centered on course content. *Understanding others* means LAs were identifying other people's perspectives. *Team work* means the LAs facilitate student group work and LAs themselves were working in a group. *General communication* means communication that was not necessarily tied to course content but represented broader communication activity such as talking in front of people, presenting slides, or dealing with individual student issues. The categories progress generally from more technical knowledge and skills to more social ones. The category of understanding others resides in both technical (as in understanding how other people think of a technical question) and social (the perspective of a professor vs. a student). The coding of categories were decided by both the words LAs said and the near context. The authors discussed code categories and reached agreement.

To answer our second research question: *How do the LAs' perceptions of learning align with their description of their roles in the course and the overall LA experience?* we looked for places when LAs brought the similar description when they were describing their understanding of (1) the *role* of an LA (yielded by Question "Can you describe your understanding of the role of a learning assistant?") and (2) the *knowledge and skills* LAs should have (yielded by Question "What knowledge and skills do you think needed to be an LA?"); and (3) their LA *experience* (yielded by three questions: "Can you share with us a story that comes to your mind about your LA experience that particularly influenced you?" "What are the successes that you have had as an LA?" and "Have you encountered any difficulties?").

## Results

Results are presented through tables that show a matrix of learning categories (Table 4) and interview question categories (Table 3). Each cell in the table describes a certain kind of learning (row) when LAs responded to a certain kind of question (column). For each table cell, we provide the number of LAs that provided a response coded in that category (out of the total number of LAs of the same LA context) and an example transcript excerpt. We provide three such tables, one for each LA context: studio, lab, and recitation. We describe each context based on LAs' responses to the *facts* questions, the researcher's field observations, and information available on the university course website.

### *Context 1: CBEE Studio (three LAs)*

CBEE 211 covers material balances, including thermophysical and thermochemical calculations, with 259 students enrolled in the class in Fall 2016. The LAs were facilitating Studios [16]. Studio sections (24- 30 students) were interspersed between lectures. In each studio, students formed small groups and did an activity in which they were required to complete a worksheet that consisted of problem scenarios that included conceptual and numerical questions designed to either reinforce content from the previous lecture or foreshadow the following lecture. Before the LA Program was introduced, the course only had one graduate TA in each section.

Starting Fall 2016, each studio had one graduate TA and one undergraduate LA. The TA and LA were expected to circulate around the room, interact with students and groups, and ask facilitative questions to help students get unstuck and promote learning. The social interaction between students themselves and the student and instructor was strongly encouraged. The instructional team (LAs, graduate TAs, and professor) met every week to go through studio worksheet. LAs could host office hours jointly with graduate TAs, depending on personal willingness and schedule. LAs also participated in homework and mid-term grading with the instructional team.

Table 5 presents the learning matrix, including the number of coded responses and sample excerpts: 37 coded responses out of a possible 60 (37/60). All the LAs expressed their learning in all the four coding categories (last column) and the majority of the LAs described their LA experiences that addressed the learning categories (second last column).

**Table 5.** Learning matrix, number of coded responses, and sample excerpts for the studio context.

|  | Role | Knowledge | Experience | Learning |
|---|---|---|---|---|
| **Understanding Content** | (0/3) | (3/3) I obviously need to know material. | (2/3) I think that being an LA made me more comfortable with the material. | (3/3) When I was reviewing the material for the class as it can help me in certain ways. |
| **Communicating disciplinary knowledge** | (2/3) To answer questions about the material. To facilitate learning. | (2/3) The different ways to get students to think more critically or answer questions. | (2/3) More comparable explaining difficult concepts to students. | (3/3) To explain the underlying concepts behind certain problems rather than just tell students directly how to do one problem. |
| **Understanding others** | (2/3) A peer-to-peer level way of contacting. To put yourself in their position. | (1/3) You can't mold them into becoming you because they are not you. | (2/3) There was a bunch of different ways to do it. | (3/3) You learn a different way to think if you explain it to someone else. |
| **Team work** | (1/3) To facilitate discussion about the material. | (1/3) To prompt discussion in groups. | (2/3) They actually sat down and made sure that everyone is on the same page and I saw a huge improvement. | (3/3) It helps work with other students, which is pretty important now. |
| **General communication** | (0/3) | (0/3) | (2/3) One day after class one student asked suggestions about how to feel confident in college. | (3/3) It got me used to going to meetings, and sit with group of people and talk about something. |

*Context 2: ECE Lab (four LAs)*

ECE111 is the introduction to the electrical and computer engineering profession. It is the first course many students take when they arrive at Oregon State University in the pre-electrical engineering program. This course covers the foundations of engineering problem solving and other skills necessary for success. Students are taught engineering practice through hands-on approaches. Labs meet every week and last two hours. This course had a history of using undergraduate TAs in labs. Introducing the LA Program added the pedagogy workshop to the LAs.

In Fall 2016, the labs practiced engineering design through the use of a newly introduced electric board and small design projects. Each lab session had 20-22 students and three LAs. LAs conducted demonstrations and gave a short presentation at the beginning of the lab and answering student questions during the lab. At the end, the LAs were responsible for checking the student work for completion. The LAs were also assigned to host office hours. One graduate TA led the LAs, meeting them every week and typically discussing the lab of next week. Because this was a new lab, the graduate TA had to re-design each lab process and write the lab instruction. LAs also participated in developing, writing, and testing of the lab processes.

Table 6 presents the learning matrix, including the number of coded responses and sample excerpts: 37 coded responses out of a possible 80 (37/80). Understanding the course content was mentioned by all four LAs both when they described the knowledge of skills they thought needed to be an LA and the things they viewed have learned from the LA experience (two cells in the first row).

**Table 6.** Learning matrix, number of coded responses, and sample excerpts for the Lab context

| | Role | Knowledge | Experience | learning |
|---|---|---|---|---|
| **Understanding content** | (2/4) Having enough the experience and knowledge of a specific subject to able to help people fully. | (4/4) Another knowledge set is the knowledge of the equipment itself, which is from preparation work before the lab. | (2/4) Most of the successes are also building my own understanding of things. | (4/4) My I level of understanding some material. |
| **Communicating disciplinary knowledge** | (3/4) To guide students through the lab or through the class and through their academics | (2/4) Somebody has a question, and you have an answer with an explanation and possibly another question to get them thinking. | (4/4) That was the best experience because I'm helping them with what they don't understand directly. | (1/4) You also have to think about the best way to try and explain something to a given the student. |
| **Understanding others** | (1/4) The job is more about guiding the students | (1/4) Because everyone feels nervous. A lot | (2/4) You need to change how you are thinking about it, so that | (4/4) The students will always think differently and |
|  | through college at this stage. | of people don't realize that. | you can think about how they are thinking. | so you get to see a lot of different ways of approaching problems. |
| **Team work** | (0/4) | (0/4) | (0/4) | (1/4) Making sure we can work better as a core teaching group. |
| **General communication** | (0/4) | (3/4) Another set is social skills, knowing how to interact with people. | (1/4) Success to me would be getting the student to calm down, finding out what the problem is. | (2/4) Definitely talking to larger groups of people. |

*Context 3: MIME and ENGR Recitation (four LAs)*

MIME 101, ENGR 111, and ENGR 199 are introductory engineering courses. MIME 101 provides students with an overview of mechanical, industrial, manufacturing, and energy systems engineering careers and an introduction to technical areas of study. ENGR 111 and 199 introduce engineering as a profession, historical development, ethics, curricula and engineering careers. The courses also cover introduction to problem analysis and solution, data collection, accuracy and variability. LAs hold recitations each week. ENGR 199 recitations are one hour each. MIME 101 and ENGR 111 recitations are two-hours each. LAs did short presentations at the beginning. Students then worked on their task and LAs answered questions if students had any. In the end, LAs summarized the session. These two course also had a history of using undergraduate TAs in recitations and the LA pedagogy workshop was new. The interactional team typically met weekly.

Table 7 presents the learning matrix, including the number of coded responses and sample excerpts: 34 coded responses out of a possible 80 (34/80). Communication skills (disciplinary and general) were mentioned the most among other categories (second and last rows).

**Table 7.** Learning matrix, number of coded responses, and sample excerpts for the recitation context

|  | **Role** | **Knowledge** | **Experience** | **Learning** |
|---|---|---|---|---|
| **Understanding content** | (0/4) | (3/4) We had the class ourselves before we know we have a general | (0/4) | (2/4) That kind of throughout the technical skill thing. |
| | | and solid idea what it is about. | | |
| **Communicating disciplinary knowledge** | (4/4) To help the students understand what the material is. | (2/4) How to deal with the specific questions about asking leading questions. | (2/4) I think one of the biggest difficulties is the students want to just get you to tell them the answer. | (4/4) Every time I try to improve the explanations and discussions of the materials. |
| **Understanding others** | (0/4) | (2/4) To understand all this person's doing and maybe in a different way. | (2/4) I should ask you more in the beginning what you're specifically was struggling with. | (3/4) The working with other students and seeing the different perspectives on learning. |
| **Team work** | (0/4) | (0/4) | (0/4) | (1/4) Because then that will add more to the rocky start at the beginning with dealing with things like the group discussions I had no experience doing. |
| **General communication** | (1/4) Speaking from slides the professor prepared doing in class activities. | (0/4) | (4/4) It was really important to build relationships with all of them so they could freely talk when they had questions | (4/4) It's helping a lot in terms of public speaking, talking in front of people. |

<u>*Answer to Research Question 1:*</u> *What are the ways that LAs perceive that the Engineering LA Program helps them learn knowledge and skills that are useful in engineering practice?*

The last column of the three matrices show the LAs' perception of learning.

The tables show that all the CBEE 211 Studio LAs we interviewed saw their learning in all the five categories coded. All the ECE111 Lab LAs we interviewed saw their learning in understanding content and understanding other perspectives and part of these LAs perceived their learning of other aspects. All the MIME and ENGR recitation LAs we interviewed saw their learning in communicating disciplinary knowledge and general communication and part of them saw their learning in other aspects.

The differences may be a result of the LA course context, the expectation of the LAs, and their interactions with the instructional team. For example, the Labs and Recitations were not set up to emphasize student group work and few LAs in those setting mentioned learning about team work. Whereas the Studios were designed to directly encourage student group interactions and all the three LAs mentioned learning about team work. Also, the CBEE course is more technically focused than the MIME and ENGR general introductory courses. All the LAs in the CBEE course mentioned strengthened understanding of the course content for themselves, whereas the LAs in the general introductory courses did not all identify that aspect. Although the ECE course was also introductory, the lab materials did include technical skills such as connecting electric board and programming the behaviors of circuits. In addition, the lab introduced a new board that term and all four LAs in the ECE course mentioned learning about the course content. Across contexts, learning about understanding other perspectives was brought up by almost all the LAs (10/11).

***Answer to Research Question 2:*** *How do the LAs' perceptions of learning align with their description of their roles in the course and the overall LA experience?*

We found general patterns across context and special patterns in particular LA contexts.

Roles: Nine LAs mentioned communicating disciplinary knowledge in their understanding of the role of an LA. Two out of three LAs in the CBEE Studios also said that the LAs were in a position of better understanding other students' perspective because they had taken the same course before (unlike many of the graduate TAs who tended to come from other universities). Other aspects were not emphasized by the majority of the LAs when talking about their understanding of their roles.

Knowledge and skills: When talking about knowledge and skills needed to be an LA, all 11 LAs mentioned understanding course content and six LAs mentioned communication of content knowledge. Three LAs (all of whom were ECE Lab LAs) mentioned general communication skills. Only one LA mentioned facilitating group discussion and the LA was in CBEE Studios.

Experience: The CBEE Studio LAs described the five aspects of learning evenly in their LA experience. Two out of three LAs mentioned each category. All the ECE Lab LAs described the communication of disciplinary knowledge in their experience and none of them brought up group work. All the Recitation LAs mentioned general communication and none of them mentioned group work nor content knowledge. This distribution makes sense because Recitation LAs were in the general introductory courses that had less technical content. Neither the ECE Labs nor the introductory course recitations emphasized student group work.

## Discussion

In this paper, we described an adapted engineering implementation of the LA Program. We analyzed interviews with 11 LAs from three different contexts (studio, lab, recitation) in five courses. Our analysis shows that the LAs perceived that they learned a wide array of socio-technical knowledge and skills, including course content knowledge, communication of disciplinary knowledge, understanding other perspectives, team work, and general communication. LAs in different teaching and facilitating contexts show different focus when describing their learning, roles, and experiences.

This study has several limitations. The number of LAs we interviewed was a small portion of the LAs that participated in the LA Program in Fall 2016: 11 out of the 50 pedagogically trained LAs (and 61 practicing LAs in total); thus, the results and interpretation only is representative of a subset of LAs in the first term of implementation in engineering at one institution. It would be useful to see how these perceptions compare as the program matures and with engineering LAs at other institutions with similar programs. The interview data were self-reports and the outcomes themselves were not measured. What the LAs described in their interview may also be only part of their perception, the part that occurred to them when triggered by the question in the interview context. Their responses do not necessarily speak to their full understanding of the experience. However, what the LAs said in the interview was still very important and informative. Their responses could indicate aspects that are more obvious and recursive to them.

Table 8 shows the cumulative results for all 11 LAs. It is encouraging that none of the cells is zero. Among the 11 LAs, at least one identified they had learned something through each of the question categories. Especially encouraging are the cells that show more than half to almost all LAs stating the learning. Examples include the expectation to understand content (10) and their reflection of actually learning it better (9); the identification of learning communication of disciplinary knowledge across different question categories (6-9); the understanding of other people's perspectives (10: cell understanding others & learning), and well-perceived learning and experience of learning general communication (7 and 9). All these aspects represent knowledge and skills that are directly related to the broader socio-technical skills engineers use in practice [9].

**Table 8.** Overall number of responses identified (out of 11 LAs)

|  | **Role** | **Knowledge** | **Experience** | **Learning** |
|---|---|---|---|---|
| **Understanding content** | 2 | 10 | 5 | 9 |
| **Communicating disciplinary knowledge** | 9 | 6 | 8 | 8 |
| **Understanding others** | 3 | 2 | 6 | 10 |
| **Team work** | 1 | 1 | 2 | 5 |

| General communication | 1 | 3 | 7 | 9 |
|---|---|---|---|---|

There is also information about ways to improve the LA Program and the implementation of evidence-based instructional practices in the courses therein. The most obvious is attending to the role of team work during learning as that was consistently identified least across all question categories. Including cooperative learning in classroom setting (such as studios) appears an effective way to develop team work skills both for the students in studios [16] and for the LAs facilitating studios.

The results of this study primarily drew data from LAs' responses in the interviews. However, their responses reflected the course structure, instructor expectations of the LAs, the ways how the instructional team interacted, and the students' learning experience in the classes LAs taught. All these factors may have affected LAs' experience and learning and that of the students with whom the LAs interact. The primary motivation for developing the LA Program was to enhance student learning in large-enrollment STEM courses. This study suggests an important additional benefit. In their role, Learning Assistants become facilitators of undergraduate students as they engage in disciplinary concepts, ideas, and practices. While the data presented here is preliminary, it suggests that by working with others (i.e., undergraduate students in the class, other LAs, graduate TAs and faculty on the instructional team), the LAs develop a broad set of socio-technical competencies that may help better prepare them for engineering practice.

## Acknowledgement

The authors are grateful for support provided by the National Science Foundation grant DUE 1347817. Any opinions, findings, and conclusions or recommendations expressed in this material are those of the authors and do not necessarily reflect the views of the National Science Foundation.## References

[1] S. Olson, and D. G. Riorda, "Engage to Excel: Producing One Million Additional College Graduates with Degrees in Science, Technology, Engineering, and Mathematics. Report to the President," *Executive Office of the President,* 2012.
[2] S. Freeman, S. L. Eddy, M. McDonough, M. K. Smith, N. Okoroafor, H. Jordt, and M. P. Wenderoth, M, P, "Active learning increases student performance in science, engineering, and mathematics," *Proceedings of the National Academy of Sciences*, vol. 111, no. 23, pp. 8410-8415, 2014.
[3] R. R. Hake, "Interactive-engagement versus traditional methods: A six-thousand-student survey of mechanics test data for introductory physics courses," *American journal of Physics*, vol. 66, no. 1, pp. 64-74, 1998.
[4] M. Prince, "Does active learning work? A review of the research," *Journal of Engineering Education*, vol. 93, no. 3, pp. 223-231, 2004.

**Appendix. Interview protocol**

1. Introduction and thanks for participation.
2. A little bit about yourself.
3. Please describe your understanding of the role of a learning assistant.
4. Where does your understanding of this role come from?
5. How did you hear about the LA position?
6. What course(s) are you in as a learning assistant?
7. Why did you want to be an LA in this course? (motivation)
8. What (classroom) activities did you participate in? What were your roles in those activities?
9. Describe the ways you were prepared for the course. (follow-up describe the Prep/team meetings; How do you participate in the pedagogical training?)
10. Can you share with us one story that comes to your mind about your experience as an LA that particularly influenced you?

11. What makes you feel this story is special?
12. Before serving as a LA, what, if any, previous experience did you have with instruction or teaching?
13. Describe your first day in the classroom/lab.
14. Did you feel prepared going in?
15. What were the successes you have had as an LA?
16. Did you encounter any difficulties as an LA?
17. How did you deal with them?
18. What knowledge and skills did you need for your work as an LA? (If the answer was too broad, follow up: can you give us some specific examples?) Where did you learn them?
19. Has working as an LA changed your own thinking about teaching and learning? (If yes) In what way? Can you give examples how it has changed your learning habits?
20. How do you think you have enhanced student group work (discussion, group project, etc.) in the course?
21. How confident do you feel facilitating class/lab as a LA? Did that change through the course?
22. How do you think your experience as an LA will help you as a student? Why?
23. How do you think your experience as an LA will help you in the future? Why?
24. What do you wish you would have known early on in your role as an LA?
25. What will you do differently next time you work as an LA?
26. What do you think you have learned from the LA experience that you wouldn't be able to learn without this experience?
27. Are there other thoughts or comments you would like to share with us?